\documentclass[showpacs,pra,twocolumn]{revtex4}

\usepackage{epsfig,amsmath}
\usepackage{subfigure}
\usepackage{graphicx}
\usepackage{dcolumn}
\usepackage{stmaryrd}
\usepackage{mathrsfs}
\usepackage{pifont}
\usepackage{amsthm}
\usepackage{amssymb}
\usepackage{bm}
\usepackage{latexsym}
\usepackage{hyperref}
\usepackage{color}

\newcommand{\beq}{\begin{equation}}
\newcommand{\eeq}{\end{equation}}
\newcommand{\beqa}{\begin{eqnarray}}
\newcommand{\eeqa}{\end{eqnarray}}

\begin{document}

\title{Effects of non-Markovian squeezed bath on the dynamics of open systems}
\date{\today }

\begin{abstract}
{Control of the dynamics of an open quantum system is crucial in quantum information processing. Basically there are two ways: one is the control on the system and the other is tuning the bath parameters. In this paper, we use the latter to analyze the non-Markovian dynamics of the open system. The model is that the system is immersed in non-Markovian squeezed baths. For the dynamics, a non-Markovian master eqation is obtained using the quantum state diffusion (QSD) equation technique for the weak system-bath couplings. We use the adiabatic evolution or quantum state transmission as examples to analyze the effects of the bath parameters: non-Markovianity $\gamma$, the squeezed direction $\theta$ and squeezed strength $r$. For the adiabatic or state transmission fidelity, the calculation results show that they both can be enhanced by a smaller $\gamma$ or bigger  $p$-quadrature. Interestingly, when $0<\theta<\pi/2$, the squeezed quadrature is determined by the combination of $r$ and $\theta$, and by numerical simulation we find that the fidelity peak occurs at $r=1-2\theta/\pi$. The fidelities increase with increasing $r$ when $r\in (0,1-2\theta/\pi]$. When $\theta\ge\pi/2$, lower fidelities are obtained due to the squeezed bath. Our results show that the dynamics of the open systems can be effectively controlled by reservoir enginerring.}

\end{abstract}

\author{Arapat Ablimit$^{1}$, Feng-Hua Ren$^{2}$, Run-Hong He$^{1}$, Yang-Yang Xie$^{1} $, Zhao-Ming Wang$^{1}$\footnote{wangzhaoming@ouc.edu.cn}}
\affiliation{$^{1}$ College of Physics and Optoelectronic Engineering, Ocean University of China, Qingdao 266100,
	China \\
	$^{2}$ School of Information and Control Engineering, Qingdao University of Technology, Qingdao 266520, China}
\maketitle

\section{Introduction.}

Open quantum system consists of system and environment, its dynamics is a basic research topic in the field of quantum information and quantum computing \cite{Ralph,Reichental, Wisniacki,Eleuch,Venuti}. Normally dissipation or decoherence of the system will occur due to the  system-bath coupling, which is one of the main problems to be solved in the manipulation of quantum devices. Based on the memory effects of the environment, the dynamics of an open quantum system is defined as Markovian or non-Markovian \cite{Vega}. For the memoryless Markovian process, the dynamical  is usually described by the Lindblad master equation with semigroup property. In the non-Markovian processes where the memory effects can not be neglected, the information backflow from the environment to the system leads to the deviation of the system dynamics from that of a dynamical semigroup, and in this case Lindblad master equation fails. A lot of methods have been developed to tackle the non-Markovian dynamics, such as non-Markovian quantum jump \cite{Piilo}, Feynman-Vernon influence functional path-integral \cite{Hu1992,zhang}, the time-evolving matrix product operator \cite{Strathearn} and QSD equation approach \cite{Diosi97}. 

For a non-Markovian environment, normally the environment is assumed to be a bosonic or fermionic coherent state \cite{zhang,Luoma,Liu}. It is the smallest uncertain state in the orthogonal component, and the quantum fluctuations of the coherent state in the phase space are equal in the direction of the orthogonal components $x$ and $p$, i.e., $\Delta x^{2}=\Delta p^{2}=\hbar/2$. With the development of the reservoir engineering technology \cite{Diehl,Li,Cavaliere}, the issue of protecting quantum information through environmental actions has attracted lots of attention. Theoretical studies have shown that a bath consisting of a squeezed vacuum can partially suppress the two level atom phase decay \cite{Gardiner}. Experimentally people have used resonant fluorescence on superconducting circuits to characterize the squeezing phenomenon in a cryogenic environment \cite{Toyli}. The squeezed bosonic bath can be generated by pumping a lumped-element Josephson parametric amplifier in superconducting circuit \cite{Murch}. For a quantum squeezed state, the quantum fluctuations of an orthogonal component can be reduced below the symmetry limit $\hbar/2$ and the quantum fluctuations of its conjugate variables rise to satisfy the Heisenberg uncertainty principle $\Delta x^{2}\Delta p^{2}\geq\hbar/4$. This provides us more adjustable parameters in controlling the system by the reservoir. Link et al. demonstrated that the system-bath coupling strength can be controlled by the squeezed direction, the model is the two level atom in a non-Markovian bath \cite{Link}.

The QSD equation approach proposed by Di\'osi \emph{et al}.  \cite{Diosi97,Diosi98} has shown its unique computational characteristics in the study of non-Markovian dynamics of open quantum systems  \cite{Ren20,wang2020,wangenergy,Flannigan,Arapat,Chen,MChen}. In this paper, by using this approach we derive a group of closed equations for solving the dynamics of the system in non-Markovian squeezed baths. The baths are assumed to be bosonic with weak system-bath couplings. We use the bath correlation function to describe the effects of bath squeezing on quantum fluctuations, and give the containing historical memory effect $O$ operators. Finally, as an demonstration, we analyze two examples: the adiabatic evolution and quantum state transmission. The effects of the non-Markovianity $\gamma$, the squeezing direction $\theta$ and squeezing strength $r$ are analyzed in these two models, respectively.

This paper is organized as follows. In Sec.~\ref{2}, we brief introduce the two mode squeezed bath and derive the non-Markovian master equation using the QSD approach. In Sec.~\ref{3}, we numerically calculate the adiabatic evolution of a three level atom, and the state transmission in the Heisenberg spin chain. Finally, conclusion is presented in Sec.~\ref{4}.

\section{Non-Markovian MASTER EQUATION}\label{2}

In the open quantum system model, the total Hamiltonian can be written as
\begin{equation}
	H_{tot}=H_{s}+H_{b}+H_{int},
\end{equation}
where $H_{s}$ and $H_{b}$ are the system and bath Hamiltonian, respectively. Suppose in multi-qubit system each qubit is coupled to its own environment. For the  bosonic squeezed bath $H_{b}=\sum_{j=1}^{N} H_b^j$, $H_b^j=\sum_{k}\omega_{k}^{j}b_{k}^{j\dagger}b_{k}^{j}$ (setting $\hbar=1$). $b_{k}^{j\dagger}$, $b_{k}^{j}$ are the $k$th mode bosonic creation and annihilation operators of the $j$th bath with frequency $\omega_{k}^{j}$. The system-bath interaction $H_{int}$ in the rotation wave approximation can be written as
\begin{equation}
	H_{int}=\underset{j,k}{\sum}\left(f_{k}^{j\ast}L_{j}^{\dagger}b_{k}^{j}+f_{k}^{j}L_{j}b_{k}^{j\dagger}\right),
\end{equation}
where $L_{j}$ is the Lindblad operator characterizing the mutual couplings between the system and the $j$th reservoir. $f_{k}^{j}$ stands for the coupling strength between the system and the $k$th mode of the $j$th bath. In the interaction picture with respect to the bath $H_{b}$, the total Hamiltonian becomes
\begin{equation}
	H^I_{tot}=H_{s}+H_{b}+\sum_{k}\left(f_{k}^{j} L_{j} b_{k}^{j\dagger} e^{i \omega^{j}_{k} t}+\text {h.c.}\right).
	\label{eq3}
\end{equation}
Now the state vector of the total system satisfies the Schr$\ddot{o}$dinger equation $\frac{\partial}{\partial t}\left|\psi(t)\right\rangle=-iH^I_{tot}\left|\psi(t)\right\rangle$.
 
In this paper, we only consider the symmetric two mode squeezed state bath as in Ref.~\cite{Link}. In this case we take $L_{j}=L_{j}^{\dagger}$, and use the squeezed vacuum state $\left|\phi\right\rangle =S(\xi)\left|0\right\rangle $ to describe the initial state of the environment. Here $S(\xi)$ is the unitary two mode squeezing operator, for the $k$th mode of the $j$th bath it can be written as
\begin{equation}
	S^{j}_{k}(\xi_{j})=e^{r_{j}(\xi_{j}^{\ast}b_{k_0+k}^{j}b_{k_0-k}^{j}-\xi_{j} b_{k_0+k}^{j\dagger}b_{k_0- k}^{j\dagger})},
\end{equation}
where $r_{j}\in[0,1]$ is the squeezing strength. When $r_{j}=0$ the $j$th bath corresponds to the non-squeezed vacuum state $\left|0\right\rangle $. $\xi_{j}=e^{i\theta_{j}}$, and $\theta_{j}$ denotes the squeezing direction. When $\theta_{j}=0, (\pi)$ the $j$th bath squeezes the $p,(x)$-quadrature. $k_{0}^{j}=\omega_{0}^{j}/c$, and $\omega_{0}$ is the squeezed center frequency. $b_{k}^{j\dagger}$, $b_{k}^{j}$ and $S^{j}_{k}(\xi_{j})$ satisfy
\begin{equation}
	S^{j\dagger}_{k-k_0}(\xi_{j})b^{j}_{k}S^{j}_{k-k_0}(\xi_{j})=u_{j}b^{j}_{k}-v_{j}b_{2k_0-k}^{j\dagger},
\end{equation}
\begin{equation}
	S^{j\dagger}_{k-k_0}(\xi_{j})b^{j\dagger}_{k}S^{j}_{k-k_0}(\xi_{j})=u_{j}b^{j\dagger}_{k}-v_{j}^{*}b^{j}_{2k_0-k}.
\end{equation}
where $u_{j}=cosh(r_{j})$, $v_{j}=w_{j}\xi_{j}$, $w_{j}=sinh(r_{j})$ and $u_{j}^{2}-\left|v_{j}\right|^{2}=1$.

For the non-Markovian bosonic baths, the time-local QSD equation can be written as \cite{Diosi97,Diosi98,wang2018}
\begin{eqnarray}
	\frac{\partial}{\partial t}\left|\psi(t,z^{\ast}_t)\right\rangle&=&
	[-iH_{s}+\underset{j}{\sum}(L_{j}z^{j\ast}_t\nonumber\\&\;&
	-L_{j}\overline{O}^{j}(t,z^{j\ast}_t))]\left|\psi(t,z^{\ast}_t)\right\rangle, 
	\label{eq7}
\end{eqnarray}
where $z^{j\ast}_t$ is the stochastic environmental noise, and $\overline{O}^{j}(t,z^{j\ast}_t)=\intop_{0}^{t}ds\alpha^{j}(t,s)O^{j}(t,s,z^{j\ast}_t)$. The $O^{j}(t,s,z^{j\ast}_t)$ operator is defined as $O^{j}(t,s,z^{j\ast}_t)\left|\psi(t,z^{j\ast}_t)\right\rangle =\delta \left|\psi(t,z^{j\ast}_t)\right\rangle/ \delta z^{j*}_s$.

In the Markovian limit, the $O$ operator is the Lindblad operator $L$. Then the initial condition of the $O$ operator is $O^{j}(t,t,z^{j\ast}_t)=L_{j}$. $\alpha^{j}(t,s)$ is the bath correlation function. For the two mode symmetric squeezed bath it can be determined by
\begin{eqnarray}
\alpha^{j}(t, s)&=&\left\langle \phi\left|\left[B_{j}(t)+B_{j}^{\dagger}(t)\right]\left[B_{j}(s)+B_{j}^{\dagger}(s)\right]\right|\phi\right\rangle\nonumber\\&\;
=&\sum_{k, k^{'}} f^{j}_{k} e^{-i \omega^{j}_{k} t} f_{k^{'}}^{j*} e^{i \omega^{j}_{k^{'}} s}\left\langle \phi\left|b^{j}_{k}
 b_{k^{'}}^{j\dagger}\right| \phi\right\rangle\nonumber\\&\;
 &+\sum_{k, k^{'}} f^{j}_{k} e^{i \omega^{j}_{k} t} f_{k^{'}}^{j*} e^{i \omega^{j}_{k^{'}} s}\left\langle \phi\left|b_{k}^{j\dagger}
 b^{j}_{k^{'}}\right| \phi\right\rangle\nonumber\\&\;
 &+\sum_{k, k^{'}} f^{j}_{k} e^{-i \omega^{j}_{k} t} f_{k^{'}}^{j*} e^{i \omega^{j}_{k^{'}} s}\left\langle \phi\left|b^{j}_{k}
 b^{j}_{k^{'}}\right| \phi\right\rangle\nonumber\\&\;
 &+\sum_{k, k^{'}} f^{j}_{k} e^{-i \omega^{j}_{k} t} f_{k^{'}}^{j*} e^{-i \omega^{j}_{k^{'}} s}\left\langle \phi\left|b_{k}^{j\dagger}
 b_{k^{'}}^{j\dagger}\right| \phi\right\rangle,
	\label{eq8}	
\end{eqnarray}
where $B_{j}(t)=\sum_{k}f^{j}_{k}b^{j}_{k}e^{-i w^{j}_{k}t}$. Assume the bath spectral density has the Lorentzian form $J_{j}(\omega_{j})=\frac{\Gamma_{j}}{2} \frac{\gamma_{j}^{2}}{\gamma_{j}^{2}+\left(\omega^{j}_{0}-\omega^{j}\right)^{2}}$  when $r_{j}=0$, which corresponds to the Ornstein-Uhlenbeck type bath correlation function $\alpha^{j}_{0}(t,s)=\frac{\Gamma_{j} \gamma_{j}}{2} e^{-i \omega^{j}_{0}(t-s)-\gamma_{j}|t-s|}$. $\Gamma_{j}$ and $\gamma_{j}$ denote the system-bath coupling strength and bandwidth of the environmental spectral density, respectively. The bandwidth $\gamma_{j}$ determines the magnitude of the environmental backaction. $\gamma_{j}\rightarrow\infty$ ($\gamma_{j}\rightarrow\ 0$) corresponds to a white (colored) noise situation, and the environment reaches the Markovian (non-Markovian) limit. As in Ref.~\cite{Link}, Eq.~(\ref{eq8}) can be decomposed as $\alpha^{j}(t, s)=\alpha^{j}_{1}(t, s)+\alpha^{j}_{2}(t, s)$, its specific form is
\begin{equation}
	\alpha^{j}_{1}(t,s)=\frac{\gamma_{j}\Gamma_{j}}{2}\left(u_{j}^{2}-v_{j}u_{j}e^{-2iw^{j}_{0}s}\right)e^{-iw^{j}_{0}\left(t-s\right)-\gamma_{j}\left|t-s\right|},
\end{equation}
\begin{equation}
	\alpha^{j}_{2}(t,s)=\frac{\gamma_{j}\Gamma_{j}}{2}\left(\left|v_{j}\right|^{2}-v_{j}^{*}u_{j}e^{2iw^{j}_{0}t}\right)e^{iw^{j}_{0}\left(t-s\right)-\gamma_{j}\left|t-s\right|}.
\end{equation}

We can see that these correlation functions are not only functions of $t-s$, but also functions of $t$ and $s$, indicating that the bath is initially non-stationary. Since the two-mode squeezed state is an entangled state, the bath correlation functions $\alpha^{j}_{1}(t,s)$ and $\alpha^{j}_{2}(t,s)$ contain both the effects of each mode on the system. Now the non-Markovian QSD equation  (Eq.~(\ref{eq7})) can be written as
\begin{eqnarray}
	\frac{\partial}{\partial t}\left|\psi(t,z^{\ast}_t)\right\rangle&=&
	[-iH_{s}+\underset{j}{\sum}(L_{j}z^{j\ast}_t-L_{j}\overline{O}_{1}^{j}(t,z^{j\ast}_t)\nonumber\\&\;&
	-L_{j}\overline{O}_{2}^{j}(t,z^{j\ast}_t))]\left|\psi(t,z^{\ast}_t)\right\rangle.
		\label{eq11}	
\end{eqnarray}

In the non-Markovian QSD method \cite{finite}, the density operator of the system is obtained by taking the statistical average of all possible random realizations of the environmental noise  $z^{j\ast}_t$, i.e., $\rho_{s}=\mathcal{M}\left[P_{t}\right]$. $P_{t}=\left|\psi(t,z^{j\ast}_t)\right\rangle \left\langle \psi(t,z^{j\ast}_t)\right|$, $\mathcal{M}\left[...\right]=\int e^{-z^{j*}_t z^{j}_t}\left[...\right]d^{2}z^{j}_t$. This process encodes large environmental effects with a huge dimensional Hilbert space as random variables. Actually, taking statistical averages of random variables is equivalent to taking trajectories in a huge Hilbert space. Therefore the time-local QSD equation has its physical meaning of statistical average. The non-Markovian master equation of the system can be written as

\begin{eqnarray}
	\frac{\partial}{\partial t}\rho_{s}&=&-i\left[H_{s},\rho_{s}\right]+\underset{j}{\sum}([L_{j},\mathcal{M}[P_{t}\overline{O}_{1}^{j\dagger}(t,z^{j\ast}_t)]]\nonumber\\&\;&
	-[L_{j},\mathcal{M}[\overline{O}^{j}_{1}(t,z^{j\ast}_t)P_{t}]]+[L_{j},\mathcal{M}[P_{t}\overline{O}_{2}^{j\dagger}(t,z^{j\ast}_t)]]\nonumber\\&\;&
	-[L_{j},\mathcal{M}[\overline{O}^{j}_{2}(t,z^{j\ast}_t)P_{t}]]).
	\label{eq12}
\end{eqnarray}

In Eq.~(\ref{eq12}),  the first term is the unitary part of the reduced dynamics, it is independent of noise variable $z^{j\ast}_t$ and only ruled by the system Hamiltonian $H_{s}$. The Lindblad operator and its associated operators $\overline{O}^{j}_{1}(t,z^{j\ast}_t)$, $\overline{O}^{j}_{2}(t,z^{j\ast}_t)$ containing the memory kernel describe relaxation and decoherence in open systems with different decay modes. This equation of motion for the density matrix $\rho_{s}$ contains all the dynamics of the system $H_{s}$ in snon-Markovian baths.
However, it is a hard task to determine the $O$ operator with the noise term  $z^{j\ast}_t$, so we invoke a perturbation technique to obtain the approximate $O$ operators  \cite{Yu99},
\begin{eqnarray}
	O^{j}(t, s, z^{\ast}_t)&=& O^{j}_{0}(t, s)+\int_{0}^{t} O^{j}_{1}(t, s, s_{1}) z_{s_{1}}^{j*} ds_{1} \nonumber\\&\;&
	+\int_{0}^{t} \int_{0}^{t} O^{j}_{2}\left(t, s, s_{1}, s_{2}\right) z_{s_{1}}^{j*} z_{s_{2}}^{j*} ds_{1} ds_{2}+\cdots \nonumber\\&\;&
	+\int_{0}^{t} \cdots \int_{0}^{t} O^{j}_{n}\left(t, s, s_{1}, \ldots, s_{n}\right) \nonumber\\&\;&
	\times z_{s_{1}}^{j*} \cdots z_{s_{n}}^{j*} ds_{1} \cdots ds_{n}+\cdots.
\end{eqnarray}

 When the system is weakly coupled to the bath, we can take an approximation $O^{j}(t,s,z^{j\ast}_t)$=$O^{j}(t,s)$ \cite{shiwf,zhao2}, i.e., $\mathcal{M}[P_{t}\overline{O}_{\eta}^{j\dagger}(t,z^{j\ast}_t)]=\rho_{s}\overline{O}_{\eta}^{j\dagger}(t)$($\eta=1,2$). The non-Markovian master equation in Eq.~(\ref{eq12}) takes a simple form  \cite{nie,luo,finite}
\begin{eqnarray}
	\frac{\partial}{\partial t}\rho_{s}&=&-i\left[H_{s},\rho_{s}\right]+\underset{j}{\sum}([L_{j},\rho_{s}\overline{O}_{1}^{j\dagger}(t)]\nonumber\\&\;&
	-[L_{j},\overline{O}^{j}_{1}(t)\rho_{s}]+[L_{j},\rho_{s}\overline{O}_{2}^{j\dagger}(t)]\nonumber\\&\;&
	-[L_{j},\overline{O}^{j}_{2}(t)\rho_{s}]).
	\label{eq14}
\end{eqnarray}

The time dependent operators $\overline{O}^{j}_{1}(t)$, $\overline{O}^{j}_{2}(t)$ can be numerically calculated by the following closed equations 
\begin{eqnarray}
	\frac{\partial}{\partial t}\overline{O}^{j}_{1}(t)&=& \alpha^{j}_{1}(0,0)L_{j}-(iw_{0}+\gamma_{j})\overline{O}^{j}_{1}(t)\nonumber\\&\;&
	+\underset{j}{\sum}([-iH_{s}-L_{j}\overline{O}^{j}_{1}(t),\overline{O}^{j}_{1}(t)]\nonumber\\&\;&
	-[L_{j}\overline{O}_{2}(t),\overline{O}^{j}_{1}(t)]).
	\label{eq15}
\end{eqnarray}
\begin{eqnarray}
	\frac{\partial}{\partial t}\overline{O}^{j}_{2}(t)&=& \alpha^{j}_{2}(0,0)L_{j}-(-iw_{0}+\gamma_{j})\overline{O}^{j}_{2}(t)\nonumber\\&\;&
+\underset{j}{\sum}([-iH_{s}-L_{j}\overline{O}^{j}_{1}(t),\overline{O}^{j}_{2}(t)]\nonumber\\&\;&
-[L_{j}\overline{O}_{2}(t),\overline{O}^{j}_{2}(t)]).
	\label{eq16}
\end{eqnarray}

Now by using Eqs. (\ref{eq15}) and (\ref{eq16}), the non-Markovian master equation in
Eq. (\ref{eq14}) can be numerically solved. In the Markovian limit, $\overline{O}^{j}_{1}(t)=\frac{1}{2}(u_{j}^{2}-v_{j}u_{j})L_{j}$, $\overline{O}^{j}_{2}(t)=\frac{1}{2}(\left|v_{j}\right|^{2}-v_{j}^{*}u_{j})L_{j}$. The Eq. (\ref{eq14}) reduces to the Lindblad form,

\begin{eqnarray}
	\frac{\partial}{\partial t}\rho_{s}&=&-i\left[H_{s},\rho_{s}\right]-\underset{j}{\sum}((u_{j}^{2}-v_{j}u_{j})(\frac{1}{2}\left\{L_{j}^{\dagger} L_{j}, \rho_{s}\right\}\nonumber\\&\;&
	-L_{j} \rho_{s} L_{j}^{\dagger}))-\underset{j}{\sum}((\left|v_{j}\right|^{2}-v_{j}^{*}u_{j})(\frac{1}{2}\left\{L_{j}^{\dagger} L_{j}, \rho_{s}\right\}\nonumber\\&\;&
	-L_{j} \rho_{s} L_{j}^{\dagger})).
	\label{eq160}
\end{eqnarray}

\section{Examples}\label{3}

\subsection{Adiabatic evolution}\label{A}

Adiabatic quantum computation requires that the system must be kept in its instantaneous ground state during the time evolution \cite{Farhi,Gosset,wang97,he}. When the system is immersed in bosonic baths, the adiabatic fidelity which measures the adiabaticity has been proven to decrease with increasing system-bath interaction strength and bandwidth of the bath, i.e., a strong bath will destroy the adiabaticity \cite{wang heat}. However, by tuning the environmental parameters, it shows that strong non-Markovinity and weak-system bath strength will be helpful to maintain the adiabaticity \cite{wang heat}. In this part, we will see that the adiabatic fidelity can be further boosted by reservoir enginerring. As an example, we consider a three level system immersed in a bosonic squeezed bath. Then the numbers of the baths $j=1$ in Eqs.~(\ref{eq14})-(\ref{eq160}). The time-dependent Hamiltonian of the system reads \cite{wujing}
\begin{equation}
	H_{s}=\left(1-\frac{t}{T}\right)J_{z}+\frac{t}{T}J_{x},
\end{equation}
where $J_{z}=\left|2\right\rangle \left\langle 2\right|-\left|0\right\rangle \left\langle 0\right|$, $J_{x}=\left(\left|2\right\rangle \left\langle 1\right|+\left|1\right\rangle \left\langle 2\right|+\left|1\right\rangle \left\langle 0\right|+\left|0\right\rangle \left\langle 1\right|\right)/\sqrt{2}$. $T$ denotes the total evolution time. From time $t=0$ to $t=T$, the system Hamiltonian $H_{s}$ changes from $J_{z}$ to $J_{x}$. Assume the system is initially prepared at the ground state $\left|0\right\rangle $. According to the adiabatic theorem when the total evolution time $T$ is long enough and the system is not affected by the environment, the final state will be $\left|\psi(T)\right\rangle=\left(\left|0\right\rangle -\sqrt{2}\left|1\right\rangle +\left|2\right\rangle \right)/2$  \cite{wujing}. Now we study the dynamics when this system is immersed in non-Markovian squeezed bath as in Eq.~(\ref{eq3}). In this model, the weak system-environment coupling ($\Gamma\ll1$) is also assumed. We also use the adiabatic fidelity $F(t)=\sqrt{\left\langle \psi(T)\right|\rho_{s}(t)\left|\psi(T)\right\rangle }$ to measure the adiabaticity, where $\rho_{s}(t)$ is the reduced density matrix of the system in Eq.~(\ref{eq14}) .

\begin{figure}
	(a)
	\centerline{\includegraphics[width=1\columnwidth]{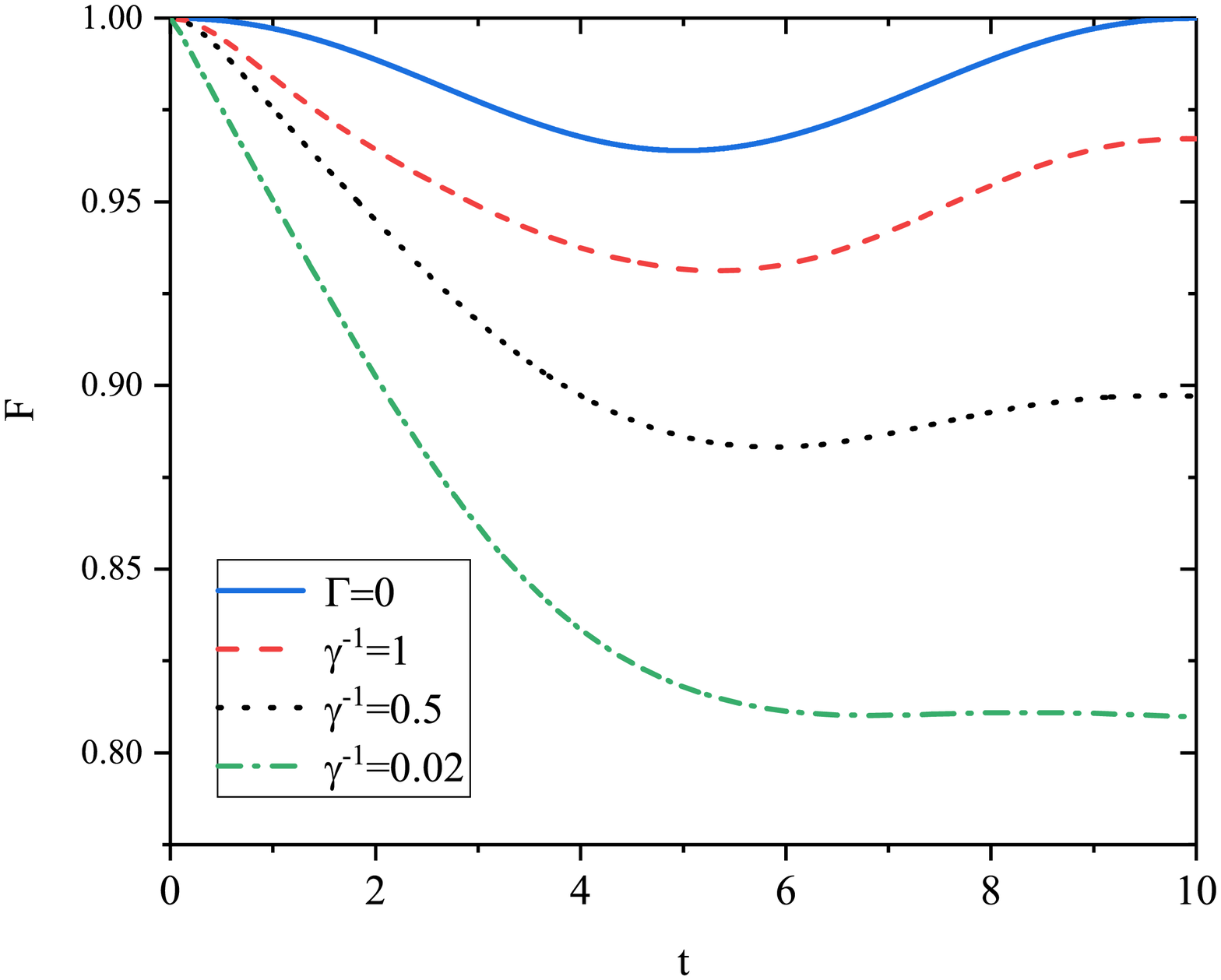}}
	(b)
	\centerline{\includegraphics[width=1\columnwidth]{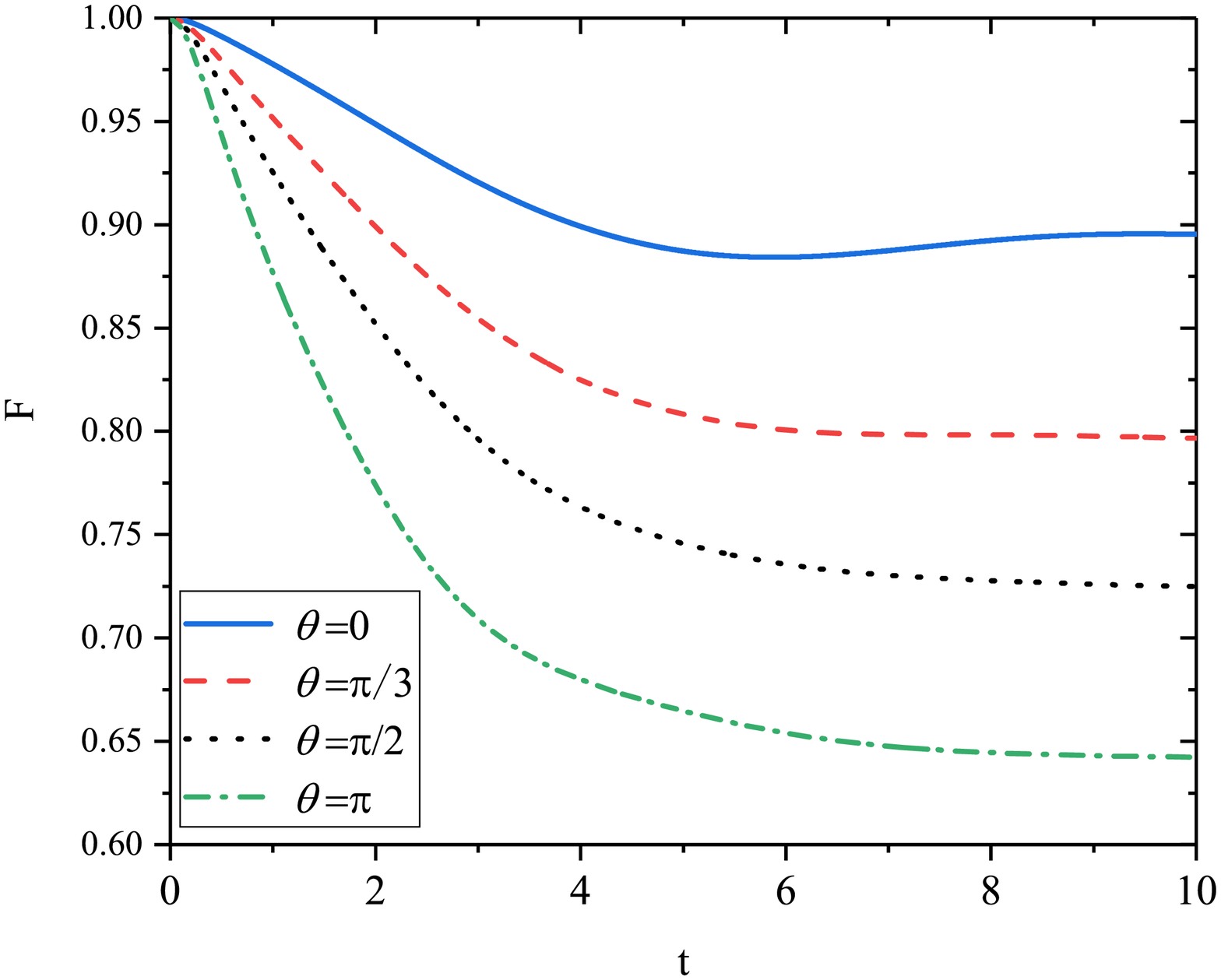}}
	\caption{(Color on line) The dynamics of the adiabatic fidelity with different environmental parameters: (a) $\gamma^{-1}$, $\theta=\pi/4$; (b) $\theta$,  $\gamma^{-1}=0.2$. Other parameters are $r=0.5$, $\Gamma=0.3$, $L=J_{x}$ and $T=10$.}
	\label{fig:1}	 
\end{figure}

In Fig.~\ref{fig:1} we plot the adiabatic fidelity $F$ versus time $t$ for different environmental  memory time $\gamma^{-1}$ (Fig.~\ref{fig:1}(a)) and squeezing parameters $\theta$ (Fig.~\ref{fig:1}(b)). The environmental parameters are taken as noise strength $\Gamma=0.3$, system-bath interaction operator $L=J_{x}$ and squeezed strength $r=0.5$. From Fig.~\ref{fig:1}(a) when there is no bath and we take the total evolution time $T=10$, $F(T)\approx 1$ which indicates that the system enters into an adiabatic regime. Clearly, the introduction of the squeezed bath will always destroy the adiabaticity. But the longer memory time (more non-Markovian) will be helpful to maintain the fidelity and this is consistent with the coherent state bath case \cite{nie}. Fig.~\ref{fig:1}(b) shows that the fidelity depends on the parameter $\theta$. $\theta=0$ corresponds to the maximum fidelity. As $\theta$ increases, i.e., the squeezed direction changes from the $p$-quadrature to the $x$-quadrature, $F$ decreases rapidly. The above results can be explained by the variances of $p$ and $x$
\begin{equation}
	V(p)=\frac{1}{4}\left[u^{2}+w^{2}-2uwcos(\theta)\right],
	\label{eq19}
\end{equation}
\begin{equation}
	V(x)=\frac{1}{4}\left[u^{2}+w^{2}+2uwcos(\theta)\right].
	\label{eq20}
\end{equation}

From Eqs.~(\ref{eq19}) and ~(\ref{eq20}), when $\theta=0$, $V(p)=e^{-2r}/4$, $V(x)=e^{2r}/4$. In this case, the quantum fluctuations of $p$-quadrature are lower than coherent states ($V(p)<1/4$), while the $x$-quadrature fluctuations are higher than coherent states ($V(x)>1/4$). The result is reversed when $\theta=\pi$, which indicates that the bath squeezing on the $p$-quadrature contributes to the adiabaticity of the system. This is due to the fact that the squeezing on the $p$-quadrature weakens the system-bath coupling. As a result, the effects of the enviromental noise on the system are attenuated \cite{Link}.

\begin{figure}[!h]
	\centerline{\includegraphics[width=1\columnwidth]{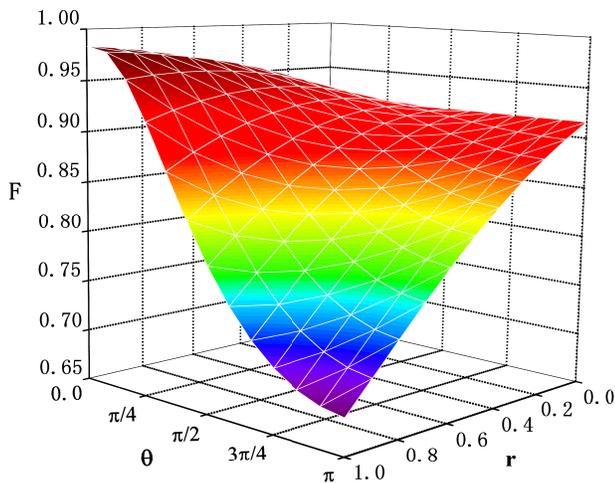}}
	\caption{(Color on line) The dynamics of the adiabatic fidelity with different squeezed  $r$ and $\theta$. Other parameters are $\gamma^{-1}=0.2$, $\Gamma=0.3$, $L=J_{x}$ and $t=1.5$. }
	\label{fig:2}	 
\end{figure}

Next we discuss the effects of squeezed strength $r$ on the adiabatic fidelity for different squeezed direction $\theta$. Here as an example, we only consider the fidelity at certain time. Fig.~\ref{fig:2} plots the fidelity versus $\theta$ and $r$ at time $t=1.5$. We can see that, when the non-Markovian bath is squeezed entirely in the $p$-quadrature ($\theta=0$, $V(p)=e^{-2r}/4$, $V(x)=e^{2r}/4$), $r$ is helpful to improve the adiabaticity, while it is squeezed in the $x$-quadrature ($\theta=\pi$, $V(p)=e^{2r}/4$, $V(x)=e^{-2r}/4$) the opposite results are obtained.

 Interestingly, when $0<\theta<\pi/2$, $F$ first increases then decreases with increasing $r$. Our numerical simulation shows that it exists a peak for certain $r$. However, when $\theta>\pi/2$, the peak disappears and $F$ always decreases with increasing $r$, which is the same as $\theta=\pi$ case. In order to understand this phenomenon, we calculate Eqs.~(\ref{eq19}) and ~(\ref{eq20}), and perform a Taylor expansion (due to $r\in[0,1]$, we only take the first three terms of the Taylor expansion), Eqs.~(\ref{eq19}) and (\ref{eq20}) can be written as  
  
 \begin{equation}
 	V(p)=\frac{1}{8}\left[2-4rcos(\theta)+4r^{2}\right],
 	\label{eq21}
 \end{equation}
 \begin{equation}
 	V(x)=\frac{1}{8}\left[2+4rcos(\theta)+4r^{2}\right].
 	\label{eq22}
 \end{equation}
 
 From Eqs.~(\ref{eq21})-(\ref{eq22}), when squeezing on both $x$ and $p$ quadrature and the squeezing on $p$ quadrature dominates ($0<\theta<\pi/2$), the variance $V(p)$ keeps decreasing as the squeezing strength $r$ increases. According to the results in Fig.~\ref{fig:1}(b), the decrease of $V(p)$ weakens the system bath coupling strength, then the fidelity is enhanced. 

\subsection{Quantum state transmission}\label{B}

Quantum state transmission through spin chains has been extensively studied for short-distance communication  \cite{wang2013,Osborne,Jayashankar}. When the communication channel is immersed in a non-squeezed environment, the transmission fidelity is found to be boosted for more non-Markovian baths and weaker system-bath interactions \cite{jpawang}. In this section, we consider the case that the baths are squeezed. As in Ref.~\cite{jpawang}, the Hamiltonian is a one dimensional $N$ site open ended $XY$ spin chain, 
\begin{equation}
	H_{s}=\sum_{i=1}^{N-1}\left[J_{i,i+1}^{x}\sigma_{i}^{x}\sigma_{i+1}^{x}+J_{i,i+1}^{y}\sigma_{i}^{y}\sigma_{i+1}^{y}\right],
\end{equation}
where $\sigma_{i}^{\alpha}(\alpha=x,y,z)$ represents the $\alpha$ component of the Pauli matrix for spins at site $i$ and $J_{i,i+1}$ is the exchange coupling strength between the neighbour sites. We take $J=J_{i,i+1}^{x}=J_{i,i+1}^{y}=-1$ which corresponds to a ferromagnetic spin chain. Assume the system is initially prepared at the $\left|\psi(0)\right\rangle=\left|100\cdots\cdots 0\right\rangle $ state. The target is to transfer the state $\left|1\right\rangle $ at one end of the chain  the other end. We still use $F(t)=\sqrt{\left\langle \psi(T)\right|\rho_{s}(t)\left|\psi(T)\right\rangle }$ to measure the transmission fidelity, where the target state is $\left|\psi(T)\right\rangle=\left|000\cdots\cdots 1\right\rangle $.
\begin{figure}[htbp]
	(a)
	\centerline{\includegraphics[width=1\columnwidth]{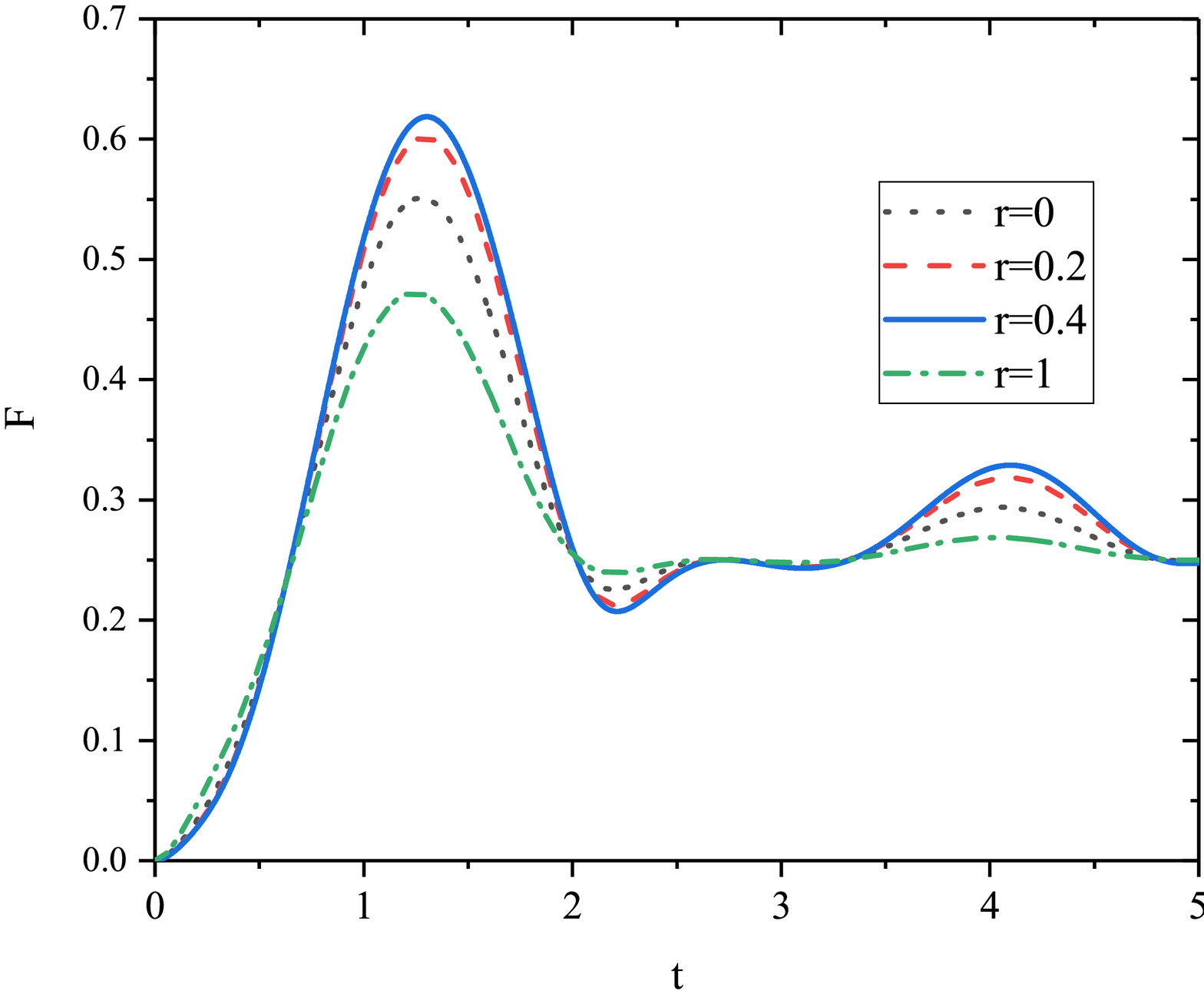}}
	(b)
	\centerline{\includegraphics[width=1\columnwidth]{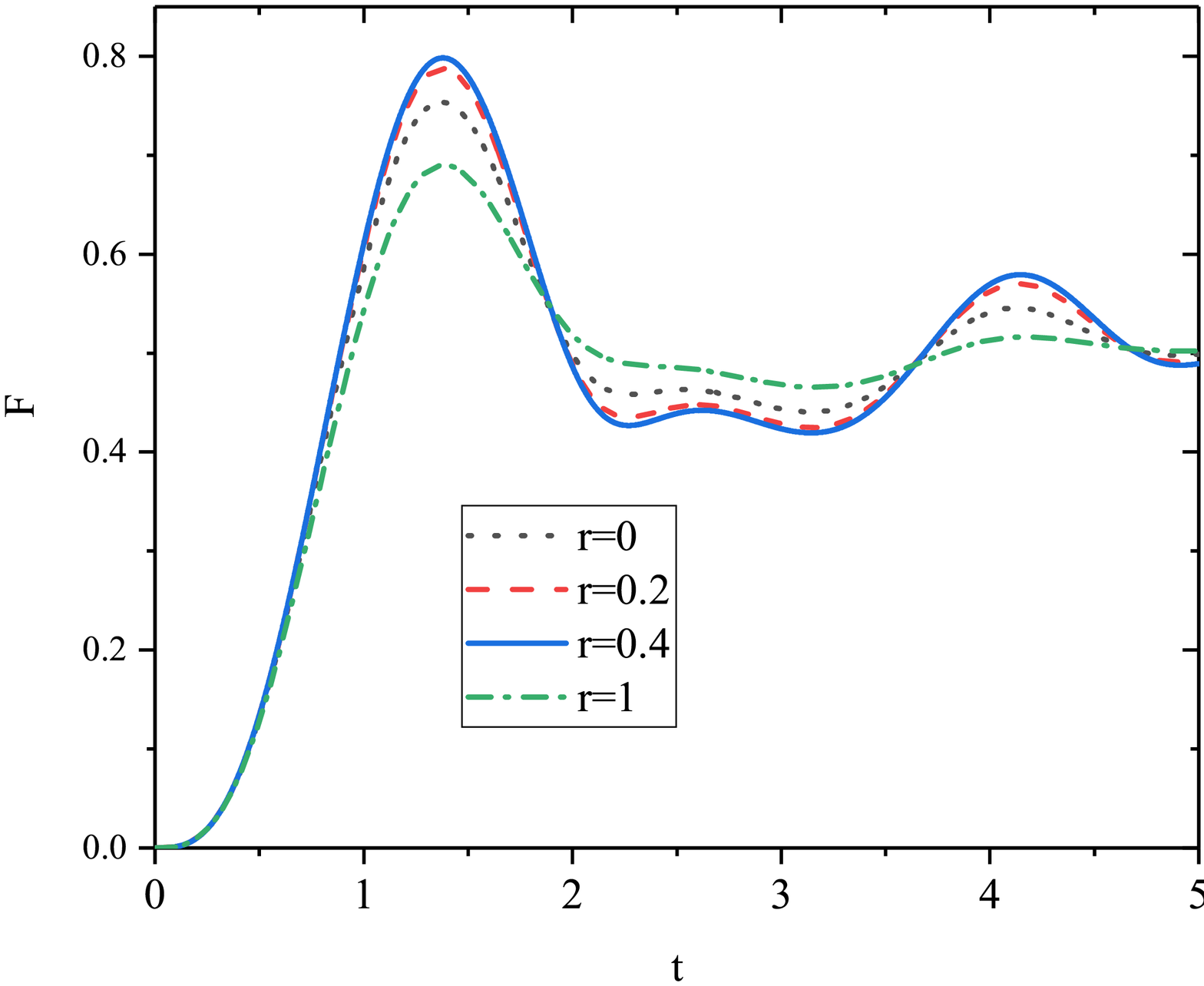}}
	(c)
	\centerline{\includegraphics[width=1\columnwidth]{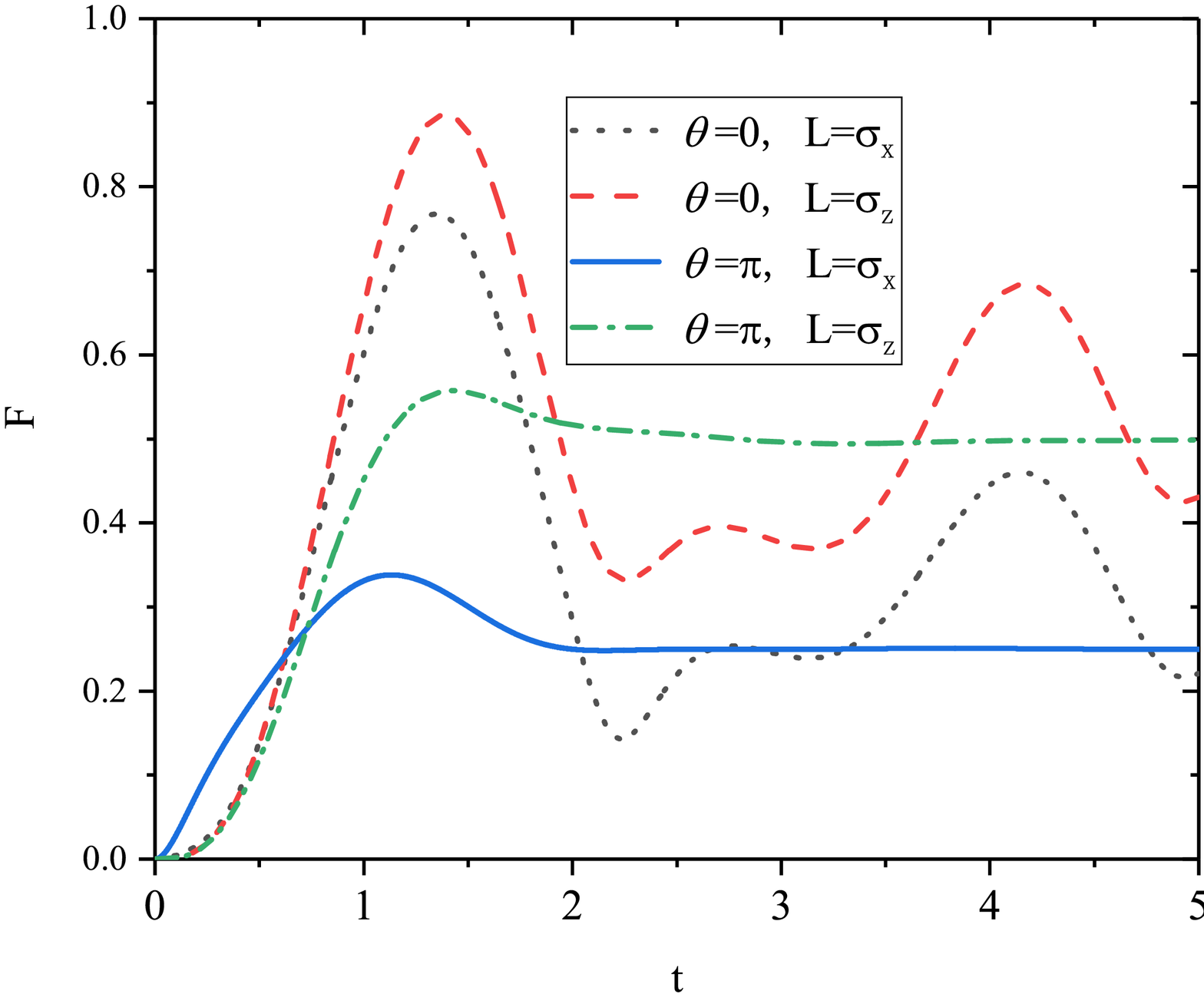}}
	\caption{(Color on line) The dynamics of the quantum state transmission fidelity with different squeezed parameters: (a) $r$, $\theta=3\pi/10$, $L=\sigma^{x}$; (b) $r$, $\theta=3\pi/10$, $L=\sigma^{z}$; (c) $r=0.5$. Other parameters are  $N=4$, $\Gamma=0.3$, $\gamma^{-1}=0.1$ and $T=5$. }
	\label{fig:3}	 
\end{figure}

In Fig.~\ref{fig:3}, we plot the state transmission fidelity dynamics for different squeezed parameters.  We consider the invidual bath model, where each spin encounters its own bath. For simplicity, assume that the neighbor spins share the same environmental parameters $\Gamma=\Gamma_{j}$, $\gamma=\gamma_{j}$, $r=r_{j}$, $\theta=\theta_{j}$,  $L=L_{j}$ for all these $j$th baths. Note that we also take the weak couplings approximations $\Gamma\ll \left|J \right| $. We compare different Lindblad operators $L=\sigma^{x}$, $L=\sigma^{z}$, and both of the two cases in Fig.~\ref{fig:3}(a), (b) and (c), respectively. The parameters are taken to be $N=4$, $\Gamma=0.3$, and $\gamma^{-1}=0.1$.  In Fig.~\ref{fig:3}(a), we take $\theta=3\pi/10$ as an example, the critical value of $r$ is $0.4$ according to the results obtained in the adiabatic model. It can be clearly seen that when $r\in(0,0.4]$, $F$ increases with increasing $r$. And after $r$ exceeds $0.4$, the opposite results are obtained. It is same to the results we obtained in the adiabatic model. Furthermore, the change in $r$ affects the non-Markovian oscillations of the system dynamics. This may be because the squeezing of the bath affects the spectrum bandwidth, and thus changes the environmental non-Markovianity \cite{Agarwal}. Fig.~\ref{fig:3} (b) takes the same parameters as Fig.~\ref{fig:3} (a) except for the dephasing. The effects of $r$ are the same as spin- boson model. Finally, a comparison between the spin boson model and the dephasing model is given in Fig.~\ref{fig:3} (c). The results show that the model performs better for the dephasing. From Fig.~\ref{fig:3}, the strong non-Markovian oscillations in the system dynamics occur, which is consistent with the effects of environmental non-Markovianity on quantum entanglement \cite{Zhao2011}. 

\begin{figure}[!h]
	\centerline{\includegraphics[width=1\columnwidth]{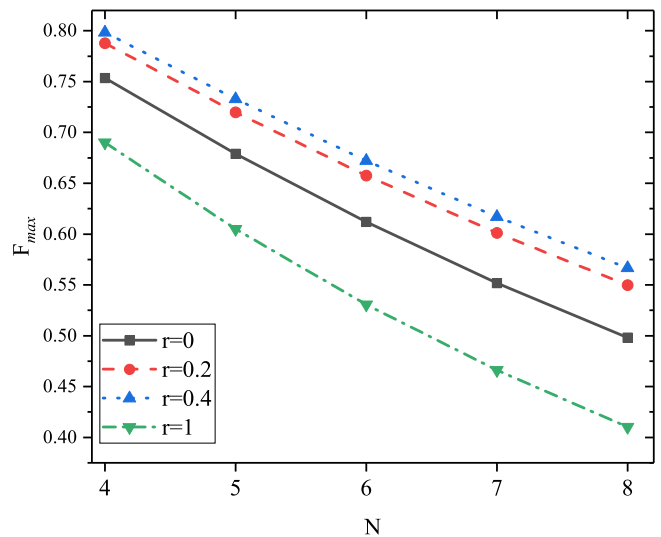}}
	\caption{(Color on line) The maximum fidelity $F_{max}$ versus length of the chain $N$ for different $r$. Other parameters are $\theta=3\pi/10$, $\Gamma=0.3$, $\gamma^{-1}=0.1$, $L=\sigma^{z}$. Clearly, the peak fidelity corresponds to $r=1-2\theta/\pi=0.4$. }
	\label{fig:4}
\end{figure}

To check if the peak fidelity observed in the adiabatic evlution is still effective in the state transmission cases for different length $N$, in Fig.~\ref{fig:4} we plot the maximum fidelity $F_{max}$ that can be achieved as a function of chain length $N$. $L=\sigma^{z}$.  
We also take $\theta=3\pi/10$, then $r=0.4$ should correspond to the peak. It is once again verified for different $N$. The peak occurs at $r=0.4$, although it decreases as $N$ increases. We conjecture that the threshold of $r=1-2\theta/\pi$ is a universal result for different system.

For the $XY$ model in the second example, the dimensionless parameters used in this paper can be converted into a dimensional form for a comparison with a recent experiment using optical lattice \cite{Jepsen}. At first, we take the couplings $J=-1,\hbar=1$ which correspond to a dimensional time variable $(-\hbar/J)t$. In the experiment, using a system of two-component bosons in an optical lattice (Bose-Hubbard model), an array of one-dimensional chains is implemented. The two states $\left\vert\uparrow\right\rangle$, $\left\vert\downarrow\right\rangle$, form a spin $1/2$ system. The effective Hamiltonian is given by a spin-1/2 Heisenberg XXZ model in the Mott insulating regime at unity filling \cite{Kuklov}, 
\begin{eqnarray}
	H=J_{xy}\sum_{i=1}^{N-1}(\sigma_i^x\sigma_{i+1}^x+\sigma_i^y\sigma_{i+1}^y)+J_z\sigma_i^z\sigma_{i+1}^z.
\end{eqnarray}

The nearest-neighbour couplings in $x,y$ are mediated by superexchange. To leading order, $J_{xy}=-4 \tilde{t}^2/U_{\uparrow \downarrow}$, $J_{z}=4 \tilde{t}^2/U_{\uparrow \downarrow}-(\tilde{t}^2/U_{\uparrow \uparrow}+\tilde{t}^2/U_{\downarrow \downarrow})$, where $\tilde{t}$ is the tunnelling matrix element between neighbouring sites, and $\tilde{t}^2/U_{\uparrow \downarrow},\tilde{t}^2/U_{\uparrow \uparrow},\tilde{t}^2/U_{\downarrow \uparrow}$ are on-site interaction energies. The couplings $J_{x,y}$ and $J_z$ can be tuned via an applied magnetic field \cite{Jepsen}. For the anisotropy $\Delta=J_z/J_{x,y}$, it can be tuned from $[-2,2]$. For our calculation, we take $J_z=0$, the evolution $t$ of up to $t=5$ ($t=5 \hbar/J_{x,y}$), which is far below the heating life time of the Mott insulator (approximately 1 $s$) \cite{Jepsen}. Also, it is in a typical time scale for nuclear magnetic resonance experiment \cite{Khurana}. In addition,
the parameters of the non-Markovian heat baths can be adjusted by reservoir engineering or simulating the influence of structured environments where the corresponding parameters can be tuned, e.g., the parameter $\gamma$ for controlled Markovian to non-Markovian transition \cite{Liu2011}, and the parameter $\Gamma$ for the detection of a weak to strong non-Markovian transition \cite{Bernardes}. The squeezed strength $r$ and direction $\theta$ of the reservoir can be experimentally controlled using the Josephson parametric amplifier pump power and phase \cite{Toyli,Murch,Flurin}.

\section{conclusions}\label{4}
In this paper, by using the QSD approach under weak coupling approximation, we obtain a non-Markovian master equation of the system in squeezed baths. Then for two examples: adiabatic evolution in a three-level system and quantum state transmission in a spin chain, we calculate the dynamics of the system. The effects of the non-Markovianity, squeezing direction and squeezing strength on the adiabatic and state transmission fidelity are analyzed, respectively. Our results show that, the two fidelities always decrease due to the presence of the environment, but they both can be enhanced for more non-Markovian baths or bigger $p$-quadrature squeezing. The $x$-quadrature squeezing leads to the enhancement of the system-bath coupling, thus decreasing the two fidelities. In addition, we find that at $0<\theta<\pi/2$, the fidelity peak appears when the squeezed strength $r$ reach a critical value $r=1-2\theta/\pi$, and its effects reverse after exceeding this critical value. We conjecture that this critical value is universal for different system models, but the only drawback is that it is obtained by numerical simulation and needs to be theoretically clarified in the following research.
By using reservoir engineering, these results can be potentially used to combat the environment noises in performing quantum information processing tasks.

\begin{acknowledgements}
We would like to thank Prof. Ahmad Abliz for the helpful discussions. This paper is supported by the Natural Science Foundation of Shandong Province (Grants No. ZR2021LLZ004).
\end{acknowledgements}

\end{document}